\documentclass[aps,prd,twocolumn,groupedaddress,showpacs,floatfix,preprintnumbers,draft]{revtex4}
\usepackage{mathrsfs}
\usepackage{amsfonts}
\usepackage{epsf} 
\begin{document}
\preprint{\tt hep-th/0505098}
\title{Moduli Stabilization and Inflation Using Wrapped Branes}
\author{Damien A. Easson}%
\email[Email:]{easson@physics.syr.edu}
\affiliation{Department of Physics, Syracuse University, Syracuse, NY 13244-1130, USA}
\author{Mark Trodden}%
\email[Email:]{trodden@physics.syr.edu}
\affiliation{Department of Physics, Syracuse University, Syracuse, NY 13244-1130, USA}
\date{\today}

\begin{abstract}
We demonstrate that a gas of wrapped branes in the early Universe can help resolve the cosmological 
Dine-Seiberg/Brustein-Steinhardt overshoot problem in the context of moduli stabilization with steep potentials in string theory. 
Starting from this mechanism, we propose a cosmological model with a natural setting in the context of an early phase 
dominated by brane and string gases. The Universe inflates at early times due to the presence of a wrapped two brane 
(domain wall) gas and all moduli are stabilized.
A natural graceful exit from the inflationary regime is achieved. However, the basic model suffers from a generalized 
domain wall/reheating problem and cannot generate a scale invariant spectrum of fluctuations without additional physics. 
Several suggestions are presented to address these issues.
\end{abstract}
\maketitle
\def\Box{\nabla^2}  
\def\ie{{\em i.e.\/}}  
\def\eg{{\em e.g.\/}}  
\def\etc{{\em etc.\/}}  
\def\etal{{\em et al.\/}}  
\def\mcS{{\mathcal S}}  
\def\I{{\mathcal I}}  
\def\mL{{\mathcal L}}  
\def\H{{\mathcal H}}  
\def\M{{\mathcal M}}  
\def\N{{\mathcal N}} 
\def\O{{\mathcal O}} 
\def\T{{\mathcal T}} 
\def\cP{{\mathcal P}} 
\def\R{{\mathcal R}}  
\def\K{{\mathcal K}}  
\def\W{{\mathcal W}} 
\def\mM{{\mathcal M}} 
\def\mJ{{\mathcal J}} 
\def\mP{{\mathbf P}} 
\def\mT{{\mathbf T}} 
\def\mR{{\mathbf R}}
\def\mS{{\mathbf S}}
\def\mX{{\mathbf X}}
\def\mZ{{\mathbf Z}}
\def\eff{{\mathrm{eff}}}  
\def\Newton{{\mathrm{Newton}}}  
\def\bulk{{\mathrm{bulk}}}  
\def\brane{{\mathrm{brane}}}  
\def\matter{{\mathrm{matter}}}  
\def\tr{{\mathrm{tr}}}  
\def\normal{{\mathrm{normal}}}  
\def\implies{\Rightarrow}  
\def\half{{1\over2}}  
\newcommand{\da}{\dot{a}}
\newcommand{\db}{\dot{b}}
\newcommand{\dn}{\dot{n}}
\newcommand{\dda}{\ddot{a}}
\newcommand{\ddb}{\ddot{b}}
\newcommand{\ddn}{\ddot{n}}
\def\be{\begin{equation}}
\def\ee{\end{equation}}
\def\bea{\begin{eqnarray}}
\def\eea{\end{eqnarray}}
\def\bs{\begin{subequations}}
\def\es{\end{subequations}}
\def\g{\gamma}
\def\G{\Gamma}
\def\vp{\varphi}
\def\mpl{M_{\rm Pl}}
\def\ms{M_{\rm s}}
\def\ls{\ell_{\rm s}}
\def\lmin{\ell_{\rm min}}
\def\lp{\ell_{\rm pl}}
\def\l{\lambda}
\def\gs{g_{\rm s}}
\def\d{\partial}
\def\co{{\cal O}}
\def\sp{\;\;\;,\;\;\;}
\def\spa{\;\;\;}
\def\r{\rho}
\def\dr{\dot r}
\def\dt{\dot\varphi}
\def\e{\epsilon}
\def\k{\kappa}
\def\m{\mu}
\def\n{\nu}
\def\om{\omega}
\def\tn{\tilde \nu}
\def\p{\phi}
\def\vp{\varphi}
\def\r{\rho}
\def\s{\sigma}
\def\t{\tau}
\def\x{\chi}
\def\z{\zeta}
\def\a{\alpha}
\def\b{\beta}
\def\de{\delta}
\def\bra#1{\left\langle #1\right|}
\def\ket#1{\left| #1\right\rangle}
\newcommand{\stt}{\small\tt}
\renewcommand{\theequation}{\arabic{section}.\arabic{equation}}
\newcommand{\eq}[1]{equation~(\ref{#1})}
\newcommand{\eqs}[2]{equations~(\ref{#1}) and~(\ref{#2})}
\newcommand{\eqto}[2]{equations~(\ref{#1}) to~(\ref{#2})}
\newcommand{\GeV}{\mbox{GeV}}
\def\kahler{K\"{a}hler\,\,}
\def\ricci{R_{\m\n} R^{\m\n}}
\def\riemann{R_{\m\n\l\s} R^{\m\n\l\s}}
\def\triemann{\tilde R_{\m\n\l\s} \tilde R^{\m\n\l\s}}
\def\tricci{\tilde R_{\m\n} \tilde R^{\m\n}}
\section{Introduction}
The problem of stabilizing string moduli (for example, the compactification volume) is a non-trivial one. This problem
is exemplified in the famous paper of Dine and Seiberg~\cite{Phys.Lett.B162.299}. In the perturbative region of moduli space,
the K\"{a}hler potential for a modulus field $\p$ typically vanishes or approaches a finite constant value in the $\p \rightarrow \infty$ limit. If the field $\p$ is the volume modulus, this corresponds to decompactification of the extra dimensions. In a time-dependent, cosmological setting, this problem is known as the Brustein-Steinhardt problem~\cite{hep-th/9212049}. Depending on the initial
conditions, the field $\p$ can acquire a large amount of kinetic energy and can easily overshoot any local minima in the potential, causing
new dimensions of space to open up. Hence, in order to create a phenomenologically acceptable cosmology, it is necessary to provide
a mechanism to lower the field gently into a local minimum of the potential, in which it can remain for an observationally acceptable length of time (i.e., longer than the Hubble time $H^{-1}$). 

In this paper, we provide such a mechanism within the context of a cosmology sourced by a gas of brane winding modes.
Such an initial state is common to the Brane Gas picture of string cosmology,
developed in~\cite{hep-th/0005212,hep-th/0109165}. 
In particular, we make use of the fact that the energy density in branes winding around compact dimensions redshifts more slowly than the kinetic energy of a typical K\"{a}hler (volume) modulus. Friction, generated by cosmic expansion, causes the modulus kinetic energy to dissipate 
rapidly~\cite{Albrecht:2001xt, Felder:2002jk}. A period of brane domination then helps the field to settle into a local minimum of the potential. Our example appears to be a simple realization
of a general phenomenon discussed in~\cite{Phys.Rev.D.58.083513,hep-ph/0010102} and more recently in~\cite{Phys.Rev.D70.126012}.

Our specific example involves a three dimensional expanding
Universe with energy density composed primarily of modulus energy, and that of string and membrane winding modes. However, the
mechanism is quite generic and may be applied to a large number of situations; for example, when a hierarchy in 
the sizes of extra-dimensions is present~\footnote{Such a hierarchy is possible in the Brane Gas picture~\cite{hep-th/0005212}.}.
We argue that this mechanism may be useful in
formulating a Brane Gas Cosmology on phenomenologically realistic compactification manifolds of non-trivial holonomy, such as Calabi-Yau spaces.

An intriguing cosmological scenario naturally arises within our basic setup. If wrapped co-dimension-one branes are present in
the early Universe, they will quickly come to dominate cosmological dynamics. In the $3+1$ dimensional expanding spacetime, these branes manifest themselves as domain walls and lead to a period of cosmic inflation~\cite{Phys.Rev.D69.083502}. The inflationary period has a natural graceful
exit mechanism and can last long enough to solve the horizon and flatness problems of the standard Big-Bang model, provided that
the ratio of the string scale to the fundamental scale is sufficiently large. 

While this is attractive, we should point out that our scenario suffers from a domain wall problem,
and is not capable, at present, of explaining the observed density fluctuations and microwave anisotropies, since the spectral index deviates significantly from the scale invariant one, $n_{s}=1$. We speculate on possible solutions to these problems in our conclusions.
\section{String Theory Background}
\begin{figure}
\begin{center}
\epsfxsize=3.5 in \centerline{\epsfbox{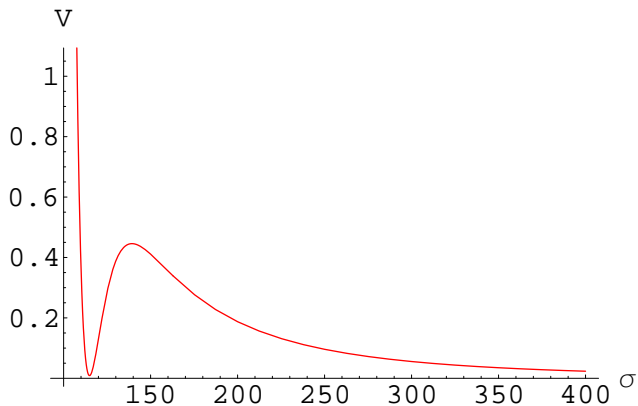}}
  \end{center}
   \caption{\label{vofs.eps} A plot of the potential (\ref{vofs}) with values $a=.1$, $A=1$, $W_{0}=-10^{-4}$
   and $D=3 \times 10^{-9}$ in units of $\mpl$.  The potential has been magnified by $5 \times 10^{-10}$.}
\end{figure}

In phenomenologically realistic string inspired models, the effective theory below the Planck
scale is typically described by a weakly-coupled, four-dimensional, $N=1$ supergravity theory. The effective action is, therefore,
\kahler invariant and described by a real \kahler  potential $K(\phi_{i})$, and a holomorphic superpotential $W(\p_{i})$, with Lagrangian density
\newpage
\bea\label{act1}
\mL &=& \sum_{i,j} \half K_{\p_{i},\bar \p_{j}} \nabla \p_{i} \nabla \bar \p_{j} \nonumber \\
&-& e^{K} \left( \sum_{i,j} K^{\p_{i},\bar \p_{j}}(D_{\p_{i}}W)(D_{\bar\p_{j}}W)^{\dagger} -3 \left|W\right|^{2} \right)
\,.
\eea
where the sum is over complex-valued, chiral superfields $\p_{i}$, and $D_{\p_{i}}=\partial_{\p_{i}} -\partial_{\p_{i}}K $.
Focusing on the case when all but one of the moduli (say $\p=\s$) are fixed (as in the model of Kachru, Kallosh, Linde and Trivedi (KKLT) \cite{Kachru:2003aw},
the potential 
\be
V = e^{K}\left[ \sum_{i,j} K^{{i},\bar {j}}(D_{{i}}W)(D_{\bar{j}}W) -3 \left|W\right|^{2} \right]
\,,
\ee
has a superpotential of the form
\be
W(\s) = W_{0} + A e^{-a\s}
\,,
\label{supv}
\ee
where $W_{0}$ is a tree-level contribution arising from fluxes and $\s$ is the single free, light modulus field. Evaluating the \kahler potential, and using (\ref{supv}) the potential becomes
\be
V(\s)=\frac{aAe^{{-a\s}}}{2\s^{2}}\left[ A e^{{-a\s}}\left( \frac{1}{3} a\s +1\right) + W_{0}\right] + \frac{D}{\s^{3}}
\,.\label{vofs}
\ee
We plot this potential in Fig.~(\ref{vofs.eps}).

\section{The Cosmology of Wrapped Brane Gases}
We are interested in studying the cosmology of a Universe filled with a gas of wrapped $p$-branes
and the energy of a single modulus field $\s$. Using the Dirac-Born-Infeld action~\cite{Boehm:2002bm}, one can derive the equation of state for a $p$-brane gas in a universe with $d$ spatial dimensions, obtaining 
\be
\mathscr{P}_{p} = \left[\frac{p+1}{d}v^{2} - \frac{p}{d} \right]\rho_{p}
\,,\label{eosp}
\ee
where $ \mathscr{P}_{p}$ is the pressure of the gas and $\rho_{p}$ is the energy density. In the relativistic limit, the branes behave as ordinary relativistic matter, with equation of state 
$\mathscr{P}_{p}\simeq\rho_{p}/d$. However, in the non-relativistic limit (wrapped branes with $|v|<<1$) the equation of state becomes $\mathscr{P}_{p}\simeq (-p/d)\rho_{p}$. 

We take as our ansatz a flat ($k=0$), $d+1$-dimensional, Friedmann-Robertson-Walker (FRW) universe, with metric 
\be\label{frw}
ds^{2} = -dt^{2}+a^{2}(t)\sum_{i=1}^{d} dx_{i}^{2}
\,,
\ee
where $a(t)$ is the scale factor. 
The resulting equations of motion are the Friedmann equation,
\be
H^{2}=\frac{16\pi G_{d+1}}{d(d-1)}\rho
\,,
\label{fried}
\ee
where $H$ is the Hubble parameter $H\equiv \dot a/a$, and the acceleration equation
\be
\frac{\ddot a}{a} = -\frac{8\pi G_{d+1}}{d(d-1)}\left[ (d-2)\rho + d\mathscr{P}\right]
\,.
\label{aeom}
\ee
Here, $\rho$ and $\mathscr{P}$ are the total energy density and pressure (wrapped branes plus modulus field) respectively, and the $(d+1)$-dimensional Newton constant is given by $G_{d+1}$. We will discuss the equation of motion for the modulus field $\s$, momentarily. 

Matter sources, obeying equation of state $\mathscr{P}=w\rho$, will redshift in the expanding space as
\be
\rho \sim a^{-d(1+w)}
\,,
\ee
so that wrapped $p$-branes with $w = -p/d$, satisfy 
\be
\rho_{wrapping}\sim a^{-(d-p)}
\,,\label{rdsft}
\ee
while the kinetic energy of the $\vp $ field obeys
\be
\rho_{\vp} \sim a^{-2d}
\,,
\ee
Hence, kinetic energy always redshifts more rapidly than other matter sources. 
With sufficient cosmic friction this dissipation of the modulus kinetic energy can allow the field to be placed gently into the shallow minimum of its potential. Of course, if other matter sources are
not present in sufficient quantities, it is possible for $\vp$ to overshoot the minimum at $\vp_{min}$ and to run off to infinity.

The authors of \cite{Phys.Rev.D70.126012}, explicitly showed that radiation can be used to slow down the modulus and 
lower it into the minimum of the potential. 
While radiation redshifts as $1/a^{4}$, from (\ref{rdsft}) we see that
string and brane winding gases redshift much more slowly, as $1/a^{2}$ and $1/a$, respectively. This means that wall and
string gases will be more efficient in stabilizing the modulus field than radiation. For radiation the scale factor evolves
as $a \propto t^{1/2}$, whereas for wrapped membranes, $a \propto t^{2}$ (see Eq.~(\ref{sfev})). Thus, the Hubble damping term in
(\ref{seom}) will be more significant for walls (and wrapped strings) than radiation. 

In the following example we take $d=3$. 
It is clear from $(\ref{act1})$, that the field $\s$ has a non-standard
kinetic term. For our purposes it is convenient to switch to a canonically normalized field $\vp $, with equation of motion
\be
\ddot \vp + dH\dot \vp +V'(\vp)=0
\,,
\label{seom}
\ee
where a prime denotes differentiation with respect to the field $\vp$. Specializing to $d=3$, we achieve this by
defining $\vp \equiv \sqrt{3/2} \ln \s$, yielding the canonical kinetic term $\half (\d \vp)^{2}$. In terms of the new variable $\vp$, the potential (\ref{supv}) then becomes
\bea
V(\vp)&=& D e^{-\sqrt{6} \vp } \nonumber \\
&+&\frac{1}{2} a e^{-a e^{\sqrt{\frac{2}{3}} \vp }-2 \sqrt{\frac{2}{3}} \vp }
 \Big(A e^{-a e^{\sqrt{\frac{2}{3}} \vp}} (1+\frac{1}{3} a e^{\sqrt{\frac{2}{3}} \vp }) \nonumber \\
 &+& W_0\Big) \,.
\label{vofp}
\eea
We work with this potential for the balance of this 
paper~\footnote{Although we have now generalized our discussion to a $d$-dimensional 
expanding spacetime, it is important to note that the specific form of the supergravity 
potential given in Eq.~(\ref{vofs}) is derived assuming $d=3$.}. Because of this, if other sources
are present in the early Universe (even in small amounts) they will quickly become dominant~\cite{Phys.Rev.D70.126012}.

\section{Cosmological Evolution}
In our scenario the Universe experiences several possible distinct epochs, depending on the initial relative energy densities of
the matter sources. The earliest epoch is potential energy dominated. The field initially has
no kinetic energy and begins falling, from rest, on a steep part of the potential. 
The value of the potential is large, and the Universe
expands rapidly. In this regime, the equations of motion may be approximated by
\bea
\dot a &=& a \sqrt{\frac{2V}{d(d-1)}}\,, \nonumber \\
\ddot \vp  &+& V'(\vp)=0
\,,\label{eominit}
\eea
where we have set $8\pi G_{d+1}=1$. For the initial conditions $\vp(t=0) = \vp_{0}$, and $\dot \vp= 0$ 
the exact solutions are
\bea
a(t) &=& \exp{\frac{1}{\sqrt{d(d-1)}} \int_{\vp_{0}}^{\vp}\frac{ \sqrt V(\vp) d\vp}{\sqrt{V(\vp_{0})- V(\vp)}}}\\
\vp(t) &=& \int_{0}^{t} dt \, \sqrt{2(V(\vp_{0})-V(\vp) }  \label{spsol}
\,.
\eea

Taking the time derivative of (\ref{spsol}), we see that if the potential is steep, $\dot \vp$ can be quite large and, therefore, the kinetic energy in the modulus can quickly become the dominant energy component.
Because of this, the Universe rapidly becomes dominated by the kinetic energy of the field $\vp$. During this epoch, the equations of motion are approximated by
\bea
\ddot \vp &+& d H \dot \vp = 0 \, \nonumber \\
\dot a &=& a \sqrt{\frac{\dot \vp^{2}}{d(d-1)}}
\,,
\eea
which can be solved exactly for $\vp$,
\be
\vp(t) -\vp' = \ln{\left( \sqrt{\frac{d}{(d-1)}} (t-t') \right)}
\,,\label{sasha}
\ee
where $\vp'$ and $t'$ are constants set by the initial conditions at the beginning of the kinetic energy dominated phase.

>From (\ref{sasha}), it is clear that the modulus field will grow large if there is nothing to stop it.
However, when brane winding modes are present, the kinetic energy, which redshifts very quickly (as $1/a^{6}$), dissipates due to the
cosmic friction generated by the presence of the string and wall gases. Assuming such sources exist in sufficient quantities,
the modulus field can now be fixed. The Universe becomes dominated by the energy of the brane sources and, depending on the relative initial densities of strings and walls, several possibilities can arise. 
If the initial energy densities of all matter components are of comparable orders of magnitude then, when kinetic energy domination ends, if walls are present, they will almost always dominate over the other components. An interesting possibility (although requiring some fine tuning) is the case in which all forms of energy densities are allowed to dominate for a short time. Since this is the most intricate possibility, we will examine such a case numerically in the following section.

\section{Numerical Results}
For the numerical analysis we wish to solve (\ref{aeom}) together with (\ref{seom}) and
we take initial conditions subject to the constraint equation (\ref{fried})~\footnote{Alternatively, we may simply solve
the Friedmann equation (\ref{fried}) together with (\ref{seom}).}. Throughout, we work in units with $\mpl^{2} = (8\pi G)^{-1} =1$,
and focus on a three-dimensional Universe ($d=3$). In this case the possible winding branes are $2$-branes and strings, for which the total energy density and pressure are given by
\bea
\rho &=& \half \dot \vp^{2} + V(\vp) + \frac{\rho^{(0)}_{p2}}{a} + \frac{\rho^{(0)}_{p1}}{a^{2}} \, \\
\mathscr{P} &=& \half \dot \vp^{2} - V(\vp) -\frac{2}{3}\frac{\rho^{(0)}_{p2}}{a} - \frac{1}{3}\frac{\rho^{(0)}_{p1}}{a^{2}}
\,,
\eea
where $\rho^{(0)}_{p1}$ and $\rho^{(0)}_{p2}$ are the initial energy densities of the wrapped strings and walls, respectively.

We will use the same parameters for the potential~(\ref{vofp}) as ~\cite{Phys.Rev.D70.126012}. The parameters are: 
$A=1.0$, ${a=0.1}$, $D=3 \times 10^{-26}$, $W_{0}=-2.96 \times 10^{13}$. For these values the potential
has a true minimum at $\s_{min}=320$, or $\vp_{min}\simeq 7.065$, with $V(\vp_{min}) \simeq 6.35 \times 10^{-35}$. The barrier separating
the minimum from asymptotic infinity is located at $\vp \simeq 7.18$ and has a height of $6.28 \times 10^{-34}$. 

For initial conditions we take $a_{0}=1$, $\vp_{0}=3$ and $\dot \vp_{0}=0$.  For the initial energy densities of string and $2$-brane winding matter we take $\r_{p1}^{(0)}=\frac{15}{2}\r_{p2}^{(0)}=5.20 \times 10^{-6} \mpl^{4}$.
These initial conditions lead to a bound example, where the brane and string sources are present in sufficient quantities to
trap the modulus field in the minimum of the potential. We present an unbound case later.

\begin{figure}
\begin{center}
\epsfxsize=3.5 in \centerline{\epsfbox{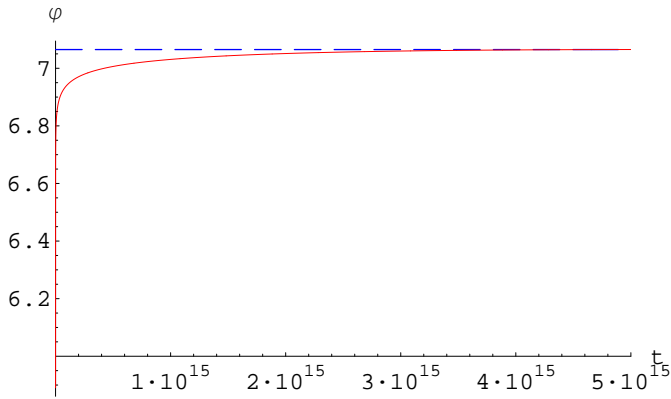}}
  \end{center}
   \caption{\label{field.eps} The time evolution of the modulus field $\vp$ (solid red line). The field becomes trapped in
   the minimum of the potential at the value $\vp_{min}=7.065$ (marked by the blue dotted line).}
\end{figure}

\begin{figure}
\begin{center}
\epsfxsize=3.5 in \centerline{\epsfbox{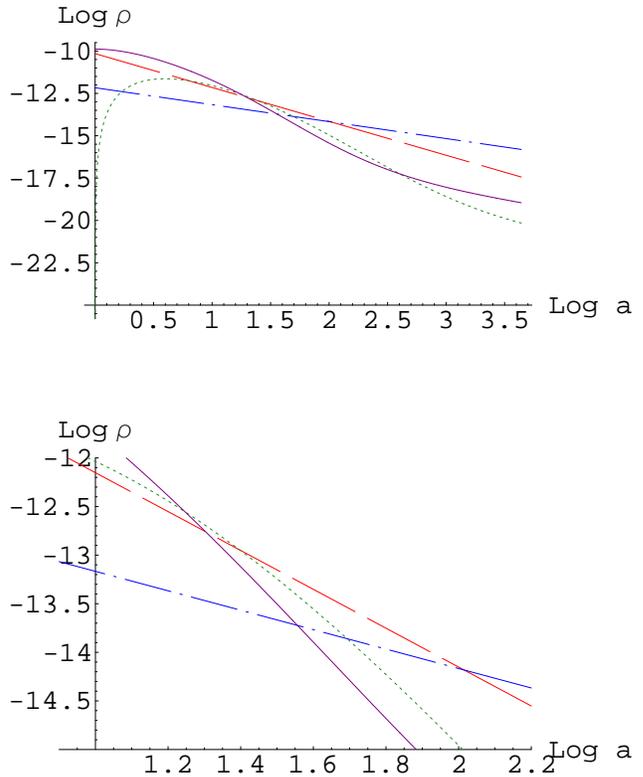}}
  \end{center}
   \caption{\label{rhovsa.eps} Top: The early time evolution of the various energy densities. The Universe starts out 
   potential dominated (solid purple line), then kinetic energy dominated (dotted green), followed by string domination (dashed red) and 
   finally wall domination (alternating dash-dot blue).
Bottom: Here we zoom in to see better the transitions from potential, to kinetic, to string and finally to wall domination.}
\end{figure}

A plot of the modulus field $\vp$ as a function of time is given in Fig.~(\ref{field.eps}).
The relative energy densities $\r_{i}$ of the moduli kinetic energy, potential, string and membrane sources are plotted as
functions of the scale factor $a$ in Fig.~(\ref{rhovsa.eps}).

Initially the Universe is dominated by potential energy, then modulus kinetic energy, followed by the string gas and then the domain walls, which redshift most slowly, as $1/a$. During wall domination the Universe inflates, since $a \sim t^{2}$. Eventually, if the walls vanish, the Universe will become potential dominated again. We will comment more on this wall inflationary epoch in the next section.

\begin{figure}
\begin{center}
\epsfxsize=3.5 in \centerline{\epsfbox{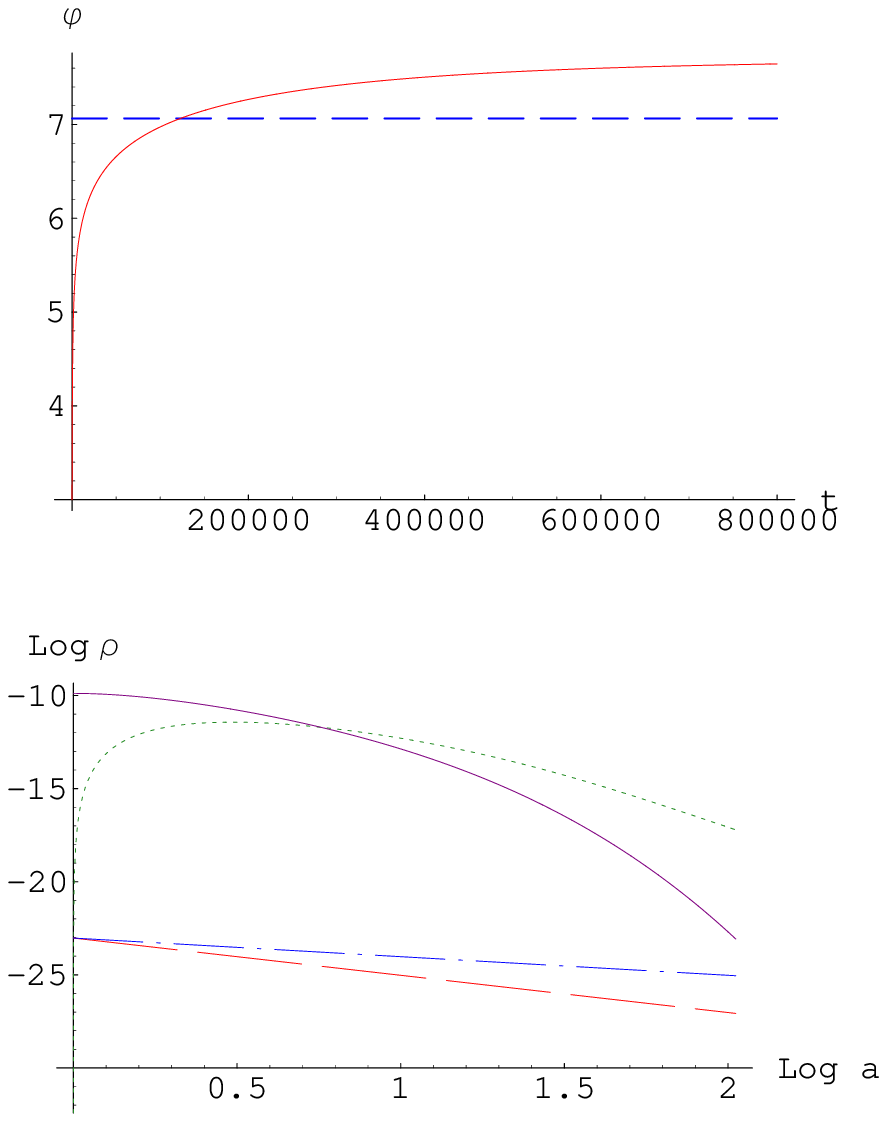}}
  \end{center}
   \caption{\label{unbd.eps} Top: Time evolution of the modulus field in the unbound case. The initial energy densities of strings and 
   branes are not sufficient to trap $\vp$. The field rolls past the local minimum of the potential, marked by the dotted line.
Bottom: The early time evolution of the various energy densities for this unbound case.}
\end{figure}

As an example of an \it unbound \rm case, we drastically dilute the energy densities of our initial sources to 
$\r_{p1}^{(0)}=\r_{p2}^{(0)}= 10^{-10} \mpl^{4}$. All other parameters are left unchanged. The resulting behavior of the modulus field and relative energy densities are plotted in Fig.~(\ref{unbd.eps}).

\section{The Inflationary Epoch}
We have argued that, in the context of our scenario, the Universe will inevitably become dominated
by a gas of wrapped membranes. In general, a gas of wrapped $p$-branes in $d$-dimensions has equation of state parameter $w=-p/d$
(recall Eq. (\ref{eosp})). Using (\ref{fried}) and (\ref{aeom}), we see that his leads to scale factor evolution of the form
\be
a(t)\propto t^{2/(d-p)} 
\,. \label{sfev}
\ee
Therefore, a gas of co-dimension-one branes ($p=d-1$)  will naturally lead to power-law 
inflation with $a(t)\sim t^{2}$, as long as the separation of the branes is much smaller
than the Hubble radius. The resulting accelerated expansion will blow up the curvature radius of the branes (relative
to the Hubble radius) and the inflationary period will end. Hence, there is no graceful exit problem in this 
scenario~\cite{Phys.Rev.D69.083502}. Specifically,
when the correlation length of the wall network $\xi(t)$ is comparable to $H^{-1}(t)=[(d-p)/2]t$, the brane gas approximation
breaks down, since the distribution of the branes averaged over a Hubble expansion time becomes inhomogeneous, and the accelerated
expansion will end. One can then show that, if the the ratio of the fundamental scale $m_{f}$ to the string scale $m_{s}$ is 
sufficiently large, $m_{f}/m_{s} \geq 10^{12}$, it is possible to obtain more than the 55-e-foldings 
required to solve the classic problems of the standard Big-Bang model~\cite{Phys.Rev.D69.083502}. 

\begin{figure}
\begin{center}
\epsfxsize=3.5 in \centerline{\epsfbox{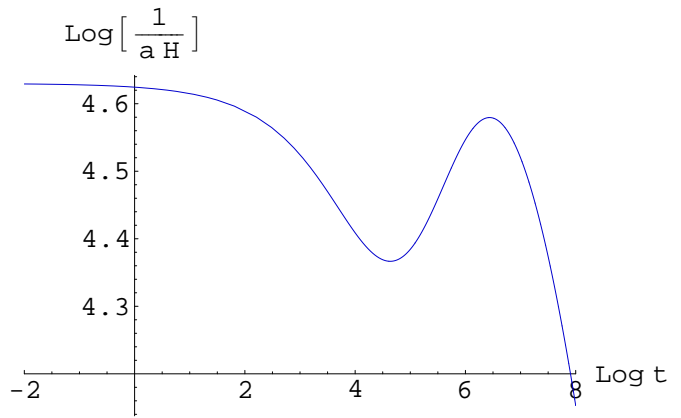}}
  \end{center}
   \caption{\label{cmh.eps} Behavior of the co-moving Hubble radius versus time. The Universe accelerates when the slope of this curve is negative.}
\end{figure}

During inflation, the co-moving Hubble length $H^{-1}/a$ is decreasing with time,
\be
\frac{d}{dt}\frac{H^{-1}}{a}<0
\,.
\ee
In our scenario, there are two periods of accelerated expansion ($\ddot a >0$). The first is during the potential dominated
phase (our initial state), during which the equations of motion are given by (\ref{eominit}). This initial inflationary period is quite
short. The modulus field $\vp$ picks up
speed, the accelerating period ends, the co-moving Hubble length increases and the Universe passes into the kinetic
energy dominated phase. The second phase of accelerated expansion is substatially longer, and begins during the 
wall dominated epoch~\footnote{A related wall inflation scenario is presented in~\cite{hep-th/0502057}}. 
This behavior of the co-moving Hubble length is clearly seen in Fig.~(\ref{cmh.eps}).

Three difficult obstacles remain. The first is a generalized domain wall problem. Since the membranes cause
the Universe to expand as $a(t)\propto t^{2}$, rather than like radiation with $a(t)\propto t^{1/2}$, a wall dominated period
would drastically affect the abundance of light elements produced during nucleosynthesis. Besides this, if today even one wall (with a tension $\geq$ the electroweak scale) were present, it would overclose the Universe. Thus, a Universe containing walls must get rid of them eventually. 

A second, possibly related problem, is how to reheat the Universe after the wall dominated period, in order to enter into a hot, radiation dominated Universe. Finally, it seems very unlikely that  a wall accelerated Universe can produce the observed density fluctuations and microwave anisotropies, since the scalar index for wall inflation is given by
\be
n_{s}-1=-\frac{2}{\a}
\,,
\ee
where $\a = 2/(d-p)$. Therefore, walls in $d=3$ give $n_{s}=0$~\cite{Phys.Rev.D69.083502}. This deviates significantly from the observed value (e.g., from WMAPext + 2dF) of $n_{s}=0.97 \pm 0.03$ ($68\%$ confidence level)~\cite{Phys.Lett.B592.1}. 

\section{Discussion}
We have shown that wrapped brane and/or string gases in the early Universe provide an efficient 
mechanism for resolving the cosmological moduli stabilization problem. Such an initial state seems quite natural in the
context of string theory and, in particular, may be relevant to the Brane Gas picture of string cosmology~\cite{hep-th/0005212}.
While some progress has been made in formulating the scenario in the context of compactification manifolds compatible with
realistic particle physics~\cite{Easson:2001fy,hep-th/0212151,Easther:2002mi}, the Brane Gas picture may greatly benefit from the mechanism suggested here~\footnote{Some interesting recent work on stabilizing moduli in the context of Brane Gas Cosmology 
can be found 
in~\cite{hep-th/0307044}, 
\cite{hep-th/0404177}, 
\cite{hep-th/0405099}, 
\cite{hep-th/0408185}, 
\cite{hep-th/0409094},
 \cite{hep-th/0409281}, 
 \cite{hep-th/0501032}, 
 \cite{hep-th/0501249}, 
\cite{hep-th/0502039}, 
\cite{hep-th/0502069}, 
\cite{hep-th/0504047},
\cite{hep-th/0504145}, 
\cite{hep-th/0504208}.}.

In addition to the stabilization of moduli, we have presented a cosmological scenario that has an attractive way of generating an inflationary epoch in the early Universe, with a natural graceful exit mechanism. This phase of accelerated expansion is generated by a gas of 
wrapped membranes. However, as mentioned in the previous section, this cosmological scenario suffers from three remaining problems.  

The first is the domain wall problem. The simplest possible solution is to 
simply not include walls in the picture in the first place. Such an initial condition is compatible with the original Brandenberger-Vafa picture
presented in~\cite{Nucl.Phys.B316.391}, in which the initial state of the Universe contains only strings. This also seems to be a natural starting point within the context of the Type IIB corner of the string moduli
space, since the IIB theory does not admit stable two-branes. Using only a gas of string winding modes will suffice to stabilize the modulus field by our mechanism. The
only downside of this picture is that a separate mechanism for inflating the Universe is needed~\footnote{Other methods of obtaining inflation in the context of
the Brane Gas picture are discussed in\cite{hep-th/0302160,hep-th/0307043,hep-th/0501194}.}. 

A second possibility is to have the walls
decay at some point after inflation. If the walls decay into radiation, this would also provide our scenario with a reheating mechanism (avoiding another obstacle). Several ways in which this can happen are discussed in~\cite{Phys.Rev.D69.083502}. We briefly summarize them here. One possibility
is that the branes are stabilized embedded branes, which decay once the temperature of the plasma drops sufficiently~\cite{Phys.Lett.B467.205,Phys.Rev.D67.043504}.
Another possibility is that the branes collide with anti-branes and annihilate, or if holes nucleate in the branes and expand at the
speed of light, disolving them~\cite{Seckel:1984ix}. Since this would occur after inflation, there would be no causal obstruction
to the local decay of sub-Hubble-volume brane networks.  A final possibility was recently proposed by Stojkovic, Freese and Starkman~\cite{hep-ph/0505026}. A sufficient abundance of primordial black holes can perforate domain walls and then the holes in the walls can grow and destroy the walls altogether. 

Another problem with our cosmological scenario is how to generate a scale invariant spectrum of density perturbations. 
This seems impossible to do within the context of wall inflation without introducing some sort of new physics. 

Finally, we note that it is possible to obtain slow-roll inflation once the potential begins to dominate (after the walls are gone). One can tune the potential so that the barrier is very flat at the top. An example of this is provided using the imaginary component (the axion) of the 
volume modulus $\vp$~\cite{hep-th/0406230}. A second condition required to build a successful model,  is that the initial conditions would have
to be tuned so that the wall and string winding gases place the modulus gently at the top of the barrier. In such a model, the modulus itself could act as the
inflaton and generate a scale invariant spectrum. We leave more detailed explorations of the above speculations to future research projects.
\medskip
\section*{Acknowledgments}
DE would like to thank T.~Biswas for informing us that related work was being done by a group at McGill and R.~Brustein for helpful discussions. 
DE thanks the organizers of the String Gas Cosmology Workshop at McGill University for 
their hospitality and for providing a stimulating research environment. 
DE and MT are supported in part by NSF-PHY-0094122 and NSF-PHY0354990, by funds from Syracuse University and by Research Corporation.

\section*{Note Added}
As this work was being written up, we became aware of related work, also almost completed. This work has since appeared~\cite{hep-th/0505151}, and contains results which have some overlap with those presented here.

\end{document}